\documentclass[twocolumn,prl]{revtex4} 
\usepackage[cp1251]{inputenc}
\usepackage[english]{babel}
\usepackage{amsmath,amssymb}
\newcommand{\beq}[1]{\begin{equation} \label{#1} }
\newcommand{\eeq}{\end{equation}}
\newcommand{\Frac}[2]{\frac
{\textstyle\lefteqn{\phantom{{}_{\mathstrut}^{\mathstrut}}} #1}
{\textstyle\lefteqn{\phantom{{}_{\mathstrut}^{\mathstrut}}} #2}}

\newcommand{\av}[1]{\langle #1 \rangle}
\begin{document}
\title{Possible stimulation of nuclear $\alpha$ decay by superfluid helium}
\author{A.~L.~Barabanov}
\affiliation{Kurchatov Institute, Moscow 123182, Russia}
\affiliation{Moscow Institute of Physics and Technology, Dolgoprudny 141700, Moscow Region, Russia}
\date{}
\begin{abstract}
It is suggested that superfluid helium (condensate of $^4$He atoms) may 
stimulate nuclear $\alpha$ decay in a situation when an $\alpha$ emitter moves 
through superfluid helium with fine-tuned velocity, so that the backward-emitted 
$\alpha$ particle is at rest in the laboratory frame. It is shown that the probability of 
stimulated $\alpha$ decay in this case may be sizable enough to be detected.
\end{abstract}
\maketitle

\section{Introduction}

Possibilities for influencing nuclear reactions by modification of the environment have been discussed many times. In particular, the highly developed field of muon catalysis belongs in this domain. The purpose of this letter is to draw attention to the possibility, in principle, of stimulating nuclear $\alpha$ decay by superfluid helium in some specific conditions.

In a Bose system the amplitude of a particle transition to a state is proportional to $\sqrt{N+1}$, where $N$ is the state occupation number. The transition probability, hence, is proportional to $N+1$ and consists of two contributions caused by stimulated and spontaneous transitions. The stimulated transitions may result in boson accumulation in a single quantum state as, e.g., it occurs for photons in a laser. Bose-Einstein condensation (it was discovered \cite{Cornell2002, Ketterle2002} in dilute gases of alkali bosonic atoms in 1995) has the same basic explanation. 

Indeed, let us show that the temperature of Bose-Einstein condensation, $T_B=(2\pi\hbar^2/m)(n/2.612)^{2/3}$ (see, e.g., \cite{Tilley1974}) results from a competition between spontaneous and stimulated transitions. Let $T$ be the temperature of a Bose gas, $V$ be the volume, $n$ be the gas density, and the population of the ground state (Bose condensate) be $N_c \simeq nV$. Spontaneous transitions cause the scattering bosons to populate the states belonging the phase space volume $4\pi Vp_T^3/3$ (the momentum $p_T$ is defined, e.g., by $p_T^2/2m=3T/2$, where $m$ is the boson mass). On the other hand, stimulated (by the condensate) transitions leave bosons in the ground state. Transitions to the ground state dominate, when:
\beq{2m}
N_c>\frac{4\pi Vp_T^3}{3(2\pi\hbar)^3}\quad\Rightarrow\quad
T<\frac{\hbar^2}{m}\left(\frac{2\pi^2n}{\sqrt{3}}\right)^{2/3}
\equiv T_0\,.
\eeq
Thus, the Bose condensate survives at low temperatures $T<T_0$. It can be seen that $T_0 $ and $T_B $ differ only by a factor of the order of unity.

Notice that the Bose-Einstein condensation occurs when the thermal de Broglie wavelength $\lambda_T=2\pi\hbar/p_T$ surpasses the interatomic distance, because Eq. (\ref{2m}) may be rewritten as:
\beq{3.2m}
n\lambda_T^3>4\pi/3.
\eeq

It is recognized nowadays that the superfluidity of liquid helium (the system of bosonic atoms $^4$He) is related to the Bose-Einstein condensation. It is seen, in particular, from the closeness of the temperature $T_B=3.1$~K (calculated for the density of liquid helium $n=2.2\cdot 10^{22}$~cm$^{-3}$) to the temperature $T_{\lambda}=2.17$~K of helium transition to superfluid state. However, owing to the interaction, the atoms of superfluid helium populate not only the lowest quantum state (as particles of an ideal Bose gas), but some set of low-lying quantum states. Both theoretical calculations and neutron scattering experiments show that at $T<1$~K approximately 10\% of the $^4$He atoms are in the lowest state (see, e.g., \cite{Bogoyavlenskii1992,Giorgini1992,Snow1995,Balibar2003} and references therein). Thus, the density of the atomic condensate can be taken to be at least $n_c=0.1n$.

Let us consider now some nuclear process implying emission of an $\alpha$ particle -- nucleus of a $^4$He atom (or $^4$He atom if the emitted $\alpha$ particle is accompanied by two electrons) with the energy varying from zero up to the value of the order of $T_{\lambda}$. For the process in superfluid helium we would expect an enhancement of the emission rate due to stimulated emission. However, it is hard to specify such a process. Indeed, typical energy release in nuclear processes is of the order of $10^6$--$10^7$~eV, whereas $T_{\lambda}\sim 10^{-4}$~eV. Let $\lambda_{\alpha}=2\pi\hbar/p_{\alpha}$ be the de Broglie wavelength of an $\alpha$ particle with the momentum $p_{\alpha}=\sqrt{2mE_{\alpha}}$ and energy $E_{\alpha}$. For $E_{\alpha}\simeq 9$~MeV we obtain $\lambda_{\alpha}\simeq 0.5\cdot 10^{-12}$~cm. Then the parameter $n_c\lambda_{\alpha}^3$ in the left-hand side of Eq. (\ref{3.2m}) is very small: $\sim 0.2\cdot 10^{-15}$.

Formally, it is possible to consider an $\alpha$ decay with the energy release $\sim 10^{-4}$~eV. But according to the Geiger--Nuttall law, the lifetime of such an $\alpha$ emitter would be of the order of $\sim 10^{10^7}$~y. Obviously, no reasonable increase of the decay rate would lower this lifetime to an observable value. Nevertheless the situation is not hopeless due to two circumstances.
\begin{enumerate}
\item After nuclear decay, an $\alpha$ particle moves with the energy $E_{\alpha}\simeq 9$~MeV (the corresponding velocity is $v_{\alpha}=\sqrt{2E_{\alpha}/m}\simeq 2\cdot 10^9$~cm/s). Let us assume that a container with superfluid helium moves as a whole with the same velocity $v_{\alpha}$ in the same direction. Then for the decaying nucleus in  the volume of helium, the $\alpha$ particle would be in the same quantum state with the momentum $p_{\alpha}=mv_{\alpha}$ as all the $N_c=n_cV$ atoms (and their nuclei -- $\alpha$ particles) of the condensate. Therefore the probability of such $\alpha$ decay would be enhanced by a factor of $N_c$. In practice, of course, it is easier to accelerate the decaying nucleus up to the velocity $v_{\alpha}$ with respect to the superfluid helium. Then the $\alpha$ particle emission strictly backwards would be stimulated.
\item The probability of spontaneous emission is proportional to the number of available states in the phase space. In the right-hand side of the first formula of Eqs. (\ref{2m}) we have taken into account all single-particle states with energy up to $p_T^2/2m$, where $p_T$ is the maximum momentum. Just for this case the momentum $p_T $ (as well as the appropriate energy $T$) has turned out very small. But the decaying nucleus emits $\alpha$ particles into a very narrow region of the momentum space. Indeed, an uncertainty of the $\alpha$ particle energy is of the order of $\hbar/\tau$ (typically much less than $T_{\lambda}\sim 10^{-4}$~eV), where $\tau$ is the nuclear lifetime (e.g. $\hbar/\tau\sim 10^{-9}$~eV when $\tau\sim 10^{-6}$~s). Thus, stimulated emission may come to light.
\end{enumerate}

Of course, there is an important question here. Is the condensate of $^4$He atoms simultaneously the condensate of their point-like nuclei -- $\alpha$ particles? To my mind, the answer is ``yes'' and it follows specifically from the results of experimental searches of the condensate in liquid helium by means of neutron inelastic scattering (see details, e.g., in \cite{Snow1995}). Indeed, neutrons interact just with the nuclei of $^4$He atoms. On the other hand, the probability of two electrons being picked up by an $\alpha$ particle (such a possibility was already noted above) could be sizable because $v_{\alpha}$ is of the same scale as the velocities of electrons of the $\alpha$ emitter. Therefore estimates for stimulated $\alpha$ decay seem worthwhile even if the condensate is formed of $^4$He atoms.

Thus, below we consider the situation of an $\alpha$ emitter moving through a superfluid helium with the velocity $v_{\alpha}$, so that the backward-emitted $\alpha$ particle is at rest with respect to the condensate. We give an estimate for the probability of stimulated $\alpha$ decay and discuss the possibility of an experimental search for this phenomenon.

\section{Probability of stimulated $\alpha$ decay}

Let us show that the probability of $\alpha$ decay really is determined by a number of available final states and this probability might be increased by a condensate. It is convenient to consider the decay caused by a time-independent perturbation $\hat V$. Then the probability of transition from an initial state $\psi_i$ with the energy $E_i$ to a final state $\psi_n$ with the energy $E_n$ during a certain time $t$ is defined by (see, e.g., \cite{Schiff1955})
\beq{12m}
P(i\to n,t)=4\,|\av{\psi_n|\hat V|\psi_i}|^2\,
\sin^2(\Delta_nt/2\hbar)/\Delta_n^2,
\eeq
where $\Delta_n=E_n-E_i$. For a fixed $t$, the right-hand side of Eq. (\ref{12m}) represents a peak with the center at $\Delta_n=0$ and the width $\Delta E\simeq\hbar/t$. Thus, if $t$ is not small, the final states belong to a narrow layer in the phase space. 

The total transition probability is determined by the sum over states within the limits of the peak or, otherwise, by the product of the right-hand side of (\ref{12m}) and the number $\Delta G(E)=\rho(E)\Delta E$ of the final states ($\rho(E)$ is the density of states). In practice one replaces the sum by the integral over $E$. Its calculation gives: $P_{if}(t)=w_{if}t$, where the transition probability per unit time is defined by Fermi's rule:
\beq{6m}
w_{if}=\frac{2\pi}{\hbar}\,|\av{\psi_f|\hat V|\psi_i}|^2\rho(E_f).
\eeq
Here $E_f=E_i$ (because $\hat V$ does not depend on $t$).

Now we notice that the initial wave function of the $\alpha$ particle, $\psi_i$, is localized inside the nucleus ($r<r_1$). The final wave function, $\psi_f$, is distributed in a macroscopic volume $V$ and, therefore, inside the nucleus takes the form: $\psi_f(r)=\psi'_f(r)\sqrt{D/V}$, where $D=\exp(-2\int_{r_1}^{r_2}\sqrt{2m(U(r)-E_{\alpha})}\,dr/\hbar)$ is the penetrability factor ($r_1$ and $r_2$ are the turning points), while $\psi'_f(r)$ is a function of the order of unity. Thus we obtain from Eq. (\ref {6m}) the usual expression for the probability of spontaneous $\alpha$ decay per unit time:
\beq{18}
w_{sp}=\frac{D}{\tau_0}\,,\quad
\frac{1}{\tau_0}\equiv
\frac{2\pi}{\hbar}
|\av{\psi'_f|\hat V|\psi_i}|^2
\frac{4\pi p^2_{\alpha}dp_{\alpha}}{(2\pi\hbar)^3dE_{\alpha}}\,.
\eeq
The quantity $1/\tau_0$ implicitly includes the factor $\Delta G$ (the number of available final states), and $\tau_0$ is of the order of the oscillation period of an $\alpha$ particle in the nucleus.

Let us now estimate the probability of stimulated $\alpha$ decay. Let $\Psi_N(1,2\ldots N)$ be the wave function of the condensate of $N$ $^4$He atoms or their nuclei -- $\alpha$ particles (a qualitative description of $\Psi_N$ has been given by Feynman \cite{Feynman1972}). The initial wave function of $N+1$ particles, normalized to unity and symmetric with respect to permutation of any two particles, looks like
\beq{18.3}
\begin{array}{l}
\Psi_i(1,2\ldots N\!+\!1)=
\Frac{1}{\sqrt{N+1}}\left[
\Psi_N(1,2\ldots N)\psi_i(N\!+\!1)\right.
\\[\bigskipamount]
\phantom{\Psi_i(1,2\ldots N\!+\!1)}
{}+\Psi_N(1\ldots N\!-\!1,N\!+\!1)\psi_i(N)+\ldots
\\[\bigskipamount]
\left.\phantom{\Psi_i(1,2\ldots N\!+\!1)}
{}+\Psi_N(2\ldots N,N\!+\!1)\psi_i(1)\right].
\end{array}
\eeq
Evidently, the matrix element over the final condensate state $\Psi_f\equiv\Psi_{N\!+\!1}$ and $\Psi_i$ given by Eq. (\ref{18.3}) is enhanced by the factor of $\sqrt{N+1}$ in comparison with the single-particle matrix element
\beq{18.4}
\av{\Psi_f|\sum_i\hat V(i)|\Psi_i}=\sqrt{N+1}\,\,
\av{\Psi_{N\!+\!1}|\hat V|\Psi_N\psi_i}.
\eeq
The probability of transition per unit time, hence, increases by the factor of $N+1$.

It should be noted that $\alpha$ decay is a tunneling process. Thus, it is instructive to show independently that the barrier penetrability for bosons increases due to symmetrization of the wave function. Before turning to this task, we note that this fact might be demonstrated experimentally by modern techniques, that allow the preparation of Bose condensates as well as single atoms in double well traps (see, e.g., \cite{Albiez2005,Shin2005,Teper2006,Sortais2007}). To do this, an initial state of a single boson in one well and Bose condensate in the other well should be formed.

Let us consider a one-dimensional double well potential, $U(x)=U(-x)$. The left and right wells are separated by a barrier. Thus, the single-particle wave function, $\psi_s(x)$, of the lowest state with the energy $E_s$ is symmetric, $\psi_s(-x)=\psi_s(x)$, whereas the single-particle wave function, $\psi_a(x)$, of the next state with the energy $E_a=E_s+\delta E$ is antisymmetric, i.e. $\psi_a(-x)=-\psi_a(x)$. Let $\psi_s (x)$ and $\psi_a(x)$ be positive at $x<0$. Then the time-dependent single-particle wave function,
\beq{101m}
\psi(x,t)=\left(
\psi_s(x)\,e^{-iE_st/\hbar}+\psi_a(x)\,e^{-iE_at/\hbar}\right)/\sqrt{2}\,,
\eeq
describes the particle localized in the left well at \mbox{$t=0$}. But at $t=\pi\hbar/\delta E$ the same function describes the particle localized in the right well. In the other words, during the time $\pi\hbar/\delta E$ the particle passes through the barrier.

Let $N$ bosons now be in the right well at $t=0$, whereas one more boson is localized in the left well. The time-dependent wave function of the whole system may be written as
\beq{102m}
\begin{array}{l}
\Psi(x_1\ldots x_{N+1},t)={}
\\[\medskipamount]
\phantom{\Psi}\left(
\Psi_s(x_1\ldots x_{N+1})\,e^{-i(N+1)E_st/\hbar}\right.
\\[\medskipamount]
\left.\phantom{\Psi}
{}+(-1)^{N}\,
\Psi_a(x_1\ldots x_{N+1})\,e^{-i(N+1)E_at/\hbar}\right)/\sqrt{2}\,,
\end{array}
\eeq
where $\Psi_s$ and $\Psi_a$ are the products of symmetric and antisymmetric single-particle functions, respectively. Assuming that at $t=0$ the coordinates $x_1$, $x_2$\ldots $x_N$ are positive (i.e., $N$ particles are in the right well), the function (\ref{102m}) is nonzero only if $x_{N+1}$ is negative (i.e. $(N+1)$th particle is in the left well). However, at $t=\pi\hbar/(N+1)\delta E$ with the same assumptions about $x_1$, $x_2$\ldots $x_N$, the wave function (\ref{102m}) describes the $(N+1)$th particle in the right well. Thus, indeed, the time of transition through the barrier decreases by the factor of $(N+1)$ (to describe accurately the initial state when $N$ bosons are in the right well, while one more boson is in the left well, one needs to introduce additional terms into the right-hand side of Eq. (\ref{102m}), but the quoted terms are leading).

Let us now estimate the ratio of the probabilities per unit time for stimulated, $w_{st}$, and spontaneous, $w_{sp}$, $\alpha$ emission. The simplest assumption is the following: among $\Delta G=\rho(E_{\alpha})\Gamma_{\alpha}$ of available final states, where the line width $\Gamma_{\alpha}=\hbar/\tau_{\alpha}$ is determined by the nuclear lifetime $\tau_{\alpha}$, there is only one state for the backward-emitted $\alpha$ particle populated by $N_c$ condensate particles. Thus, we get:
\beq{18.5}
w_{st}/w_{sp}=N_c/\Delta G.
\eeq
Taking $\tau_{\alpha}=1/(w_{sp}+w_{st})$, we obtain a quadratic equation for $w_{st}$. Its solution,
\beq{103m}
w_{st}=w_{sp}(\sqrt{1+2b}-1)/2,
\eeq
depends on the dimensionless parameter,
\beq{104m}
b=n_c\lambda_{\alpha}^2v_{\alpha}\tau_{sp},
\eeq
that is defined by the condensate density $n_c$, the wavelength $\lambda_{\alpha}=2\pi\hbar/p_{\alpha}$,  the velocity $v_{\alpha}=p_{\alpha}/m$, and the nuclear lifetime $\tau_{sp}=1/w_{sp}$ with respect to spontaneous $\alpha$ decay. 

When $b\sim 1$, the probability of stimulated $\alpha$ decay is comparable to the probability of spontaneous decay. But, curiously, it can be easily achieved: the small parameter $n_c\lambda_{\alpha}^3\sim 10^{-15}$ can be neutralized by the factor $v_{\alpha}\tau_{sp}/\lambda_{\alpha}$, that surpasses $10^{15}$ for $\tau_{sp}>10^{-6}$~s. Moreover, for relatively long-lived nuclei with lifetime $\tau_{sp}\gg 10^{-6}$~s, we get $b\gg 1 $, and therefore $w_{st}\gg w_{sp}$. However, the stimulated $\alpha$ decay may occur only during the short time $t_x=x/v_{\alpha}$ when the decaying nucleus moves through a layer of superfluid helium of thickness $x$. Thus in any case there is no reason to expect large absolute value for the probability $P_{st}=w_{st}t_x$ of stimulated decay  (we introduce also $P_{sp}=w_{sp}t_x$ as the probability of spontaneous decay during the same time $t_x$).

\section{Numerical estimates and discussion}

\begin{table}
\caption{\label{t1} The results of calculation of parameter $b$ and 
probabilities $P_{sp}$ and $P_{st}$ of spontaneous and stimulated $\alpha$ 
decay, during the flying of the nucleus through the superfluid helium, for 
several polonium and astatine isotopes; the calculations are performed for 
$n_c=0.22\cdot 10^{22}$~cm$^{-3}$ and $x=20$~cm.}
\begin{tabular}{rccccc}
\hline
& $E_{\alpha}$ (MeV) & $\tau_{sp}$ (s) & $b$ & $P_{sp}$ & $P_{st}$ \\
\hline
$^{211}$Po & 7.44 & 0.74 & $8.6\cdot 10^5$ & $1.4\cdot 10^{-8}$ & $0.9\cdot 
10^{-5}$ \\
$^{212}$Po & 8.79 & $4.3\cdot 10^{-7}$ & 0.46 & $2.3\cdot 10^{-2}$ & $4.3\cdot 
10^{-3}$ \\
$^{213}$Po & 8.38 & $0.6\cdot 10^{-5}$ & 6.6 & $1.6\cdot 10^{-3}$ & $2.3\cdot 
10^{-3}$ \\
$^{214}$Po & 7.69 & $2.4\cdot 10^{-4}$ & $2.7\cdot 10^2$ & $4.4\cdot 10^{-5}$ & 
$4.9\cdot 10^{-4}$ \\
\hline
$^{212}$At & 7.64 & 0.45 & $5.2\cdot 10^5$ & $2.3\cdot 10^{-8}$ & $1.2\cdot 
10^{-5}$ \\
$^{213}$At & 9.08 & $1.8\cdot 10^{-7}$ & 0.19 & $5.3\cdot 10^{-2}$ & $4.6\cdot 
10^{-3}$ \\
$^{214}$At & 8.81 & $0.8\cdot 10^{-6}$ & 0.85 & $1.2\cdot 10^{-2}$ & $3.9\cdot 
10^{-3}$ \\
$^{215}$At & 8.03 & $1.4\cdot 10^{-4}$ & $1.6\cdot 10^2$ & $0.7\cdot 10^{-4}$ & 
$0.6\cdot 10^{-3}$ \\
\hline
\end{tabular}
\end{table}

The value $b\simeq 1$ corresponds to the lifetime $\tau_{sp}\simeq 10^{-6}$~s. In this case $P_{st}\sim P_{sp}\sim 10^{-2}$ for realistic value $x\sim 10$~cm.

If $b\ll 1$, then
\beq{21}
w_{st}/w_{sp}\simeq b/2
\quad\Rightarrow\quad
w_{st}\simeq n_c\lambda_{\alpha}^2v_{\alpha}/2.
\eeq
Notice that $w_{st}$ does not depend on the lifetime $\tau_{sp}$ (but herewith $w_{st}\ll w_{sp}=1/\tau_{sp}$). If, e.g., $\tau_{sp}\sim 10^{-7}$~s, then $P_{sp}\sim 10^{-1}$ and $P_{st}\sim 10^{-2}$ for $x\sim 10$~cm.

At last, the case $\tau_{sp}\gg 10^{-6}$~s ($b\gg 1$) corresponds to
\beq{23}
w_{st}/w_{sp}\simeq\sqrt{b/2}
\quad\Rightarrow\quad
w_{st}\simeq\sqrt{n_c\lambda_{\alpha}^2v_{\alpha}/2\tau_{sp}}\,.
\eeq
If, e.g., $\tau_{sp}\sim 10^{-5}$~s, then $P_{st}\sim 10^{-2}$ for $x\sim 10$~cm (whereas $P_{sp}\sim 10^{-3}$).

As an illustration, the results of calculations performed for four polonium isotopes (the daughter products are lead isotopes) and four astatine isotopes (the daughter products are bismuth isotopes) are presented in the Table~\ref{t1} (the decay parameters are from \cite{Nucleonica}). For each isotope the dimensionless parameter $b$ (\ref{104m}) has been calculated for the known energy $E_{\alpha}$ (and, hence, $v_{\alpha}$ and $\lambda_{\alpha}$) and lifetime $\tau_{sp}=T_{1/2}/\ln 2$ at $n_c=0.22\cdot 10^{22}$~cm$^{-3}$. The probabilities $P_{sp}=x/(\tau_{sp}v_{\alpha})$ and $P_{st}=w_{st}x/v_{\alpha}$ of spontaneous and stimulated $\alpha$ decay are obtained for the layer thickness $x=20$~cm ($w_{st}$ is given by Eq. (\ref{103m})). We see that the probability $P_{st}$ belongs to the interval from $10^{-3}$ to $10^{-2}$ when the parameter $b$ is comparable to unity (in agreement with rough estimates presented above).

Up to now we discussed the situation with the nucleus moving through the superfluid helium with the fine-tuned velocity $v_{\alpha}$, so that the backward-emitted $\alpha$ particle is at rest in the laboratory system. However, it seems reasonable to assume that the velocity of the $\alpha$ particle can be of the order of zero-point oscillation velocity of helium atoms, $\Delta v\sim\hbar n^{1/3}/m\sim 10^3$~cm/s. The given estimate is based on the qualitative description \cite{Feynman1972} of the condensate wave function $\Psi_N$ that varies substantially on the distances $\sim n^{-1/3}$. It means that the relative uncertainty of the kinetic energy $E_M=Mv_{\alpha}^2/2$ of the $\alpha$ emitter with mass $M$ should be not worse than $\Delta E_M/E_M\sim\Delta v/v_{\alpha}\sim 10^{-6}$.

Obviously, we are dealing here with a factor that makes the detection of stimulated $\alpha$ emission difficult. The problem is not so much in the smallness of $\Delta E_M/E_M$ as in the nonconstancy of the kinetic energy of the moving nucleus in the superfluid helium (due to slowing down). One may hope, however, that this problem is solvable. Some ideas are proposed in the following, final section.

\section{Conclusion}

It is suggested that superfluid helium may influence the probability of nuclear $\alpha$ decay. Indeed, the superfluid component consists of a significant part of helium atoms and their nuclei ($\alpha$ particles) in a single quantum state (in Bose condensate). Thus, the amplitude of $\alpha$ particle emission to this state is enhanced by the factor of $\sqrt{N_c+1}$, where $N_c$ is the macroscopic occupation number.

It is assumed that the ion or neutral atom (with decaying nucleus) moves through the layer of superfluid helium, so that the backward-emitted $\alpha$ particle is at rest (with respect to the condensate of $^4$He atoms). It is shown that for $\alpha$ emitters with a lifetime $10^{-7}$--$10^{-5}$~s the probability of stimulated $\alpha$ decay during the motion of the nucleus through the layer of thickness $\sim 10$~cm may be of the order of $10^{-3}$--$10^{-2}$.

To detect the effect, it is necessary to compare probabilities of $\alpha$ decay for temperatures of helium above and below $T_{\lambda}$. The correlation would be an impressive demonstration of the possibility to influence a quantum process with characteristic energy release $\sim 10^7$~eV by a very weak impact at the level of $T_{\lambda}\sim 10^{-4}$~eV. 

Constant velocity of the moving $\alpha$ emitter (ion or atom) is necessary. Perhaps the energy losses of the ion due to slowing down can be recompensed by some additional ion acceleration directly in helium. Perhaps the ion or atom can be launched through the helium inside a nanotube without slowing down. Perhaps the ion or atom can be launched along a surface of superfluid helium, a small distance (of the order of $10^{-8}$~cm) from the surface (in a bound quantum state corresponding to the movement normal to the surface). Thus, to detect the effect, significant efforts and additional ideas may be needed.

It is useful to keep in mind that when studying new physical phenomenon something unexpected can be found. In particular, experimental study of the $\alpha$ decay of a nucleus at rest with respect to superfluid helium is of a certain interest (because the $\alpha$ particle leaving the Coulomb barrier is also ''at rest'' in some sense).

\begin{acknowledgments}
The author thanks the referees for useful comments and criticism, L.V. Grigorenko and R. Golub for helpful discussions. This work was supported in part by grants NS-3004.2008.2 and 2.1.1/4540 from MES of Russia.
\end{acknowledgments}

\end{document}